\documentclass{article}
\usepackage{amsmath, amssymb, graphicx, url}
\usepackage{psfrag}

\usepackage{blkarray}
\usepackage{multirow}
\usepackage[table]{xcolor}

\usepackage{caption}
\usepackage{subcaption}

\usepackage[customcolors]{hf-tikz}

\tikzset{style green/.style={
    set fill color=green!50!lime!60,
    set border color=white,
  },
  style cyan/.style={
    set fill color=cyan!90!blue!60,
    set border color=white,
  },
  style orange/.style={
    set fill color=orange!80!red!60,
    set border color=white,
  },
  hor/.style={
    above left offset={-0.15,0.31},
    below right offset={0.15,-0.125},
    #1
  },
  ver/.style={
    above left offset={-0.1,0.3},
    below right offset={0.15,-0.15},
    #1
  }
}

\usepackage{amsthm}
\newtheorem{problem}{Problem}
\newtheorem{observation}{Observation}
\newtheorem{lemma}[problem]{Lemma}
\newtheorem{theorem}[problem]{Theorem}

\theoremstyle{definition}
\newtheorem{definition}[problem]{Definition}
\newtheorem{example}{Example}
\newcommand{\R}[1][d]{\ensuremath{\mathbb{R}^{#1}}}
\newcommand{\proj}[1][d-1]{\ensuremath{\mathbb{P}^{#1}(\mathbb{R})}}
 \newcommand{\algesystem}{\ensuremath{(H,X)(p)}}
 \newcommand*{\Cdot}{{\scalebox{1.25}{$\,\cdot\,$}}}
    \newcommand{\framework}[1][p]{\ensuremath{(H,X,#1)}}

\title{Combinatorial rigidity and independence of generalized pinned subspace-incidence constraint systems}
\author{Menghan Wang \and Meera Sitharam }

\begin{document}

\maketitle

\begin{abstract}

Given a hypergraph $H$ with $m$ hyperedges and a set $X$ of $m$ \emph{pins},
i.e.\ globally fixed subspaces in Euclidean space $\R$,
a \emph{pinned subspace-incidence system}  is the pair $(H, X)$,
with the constraint that each pin in $X$ lies on the subspace spanned
by the point realizations in $\mathbb{R}^d$
of vertices of the corresponding hyperedge of $H$.
We are interested in combinatorial characterization of
pinned subspace-incidence systems that are \emph{minimally rigid},
i.e.\ those systems that are guaranteed to generically yield a locally
unique realization. As is customary, this is accompanied by a
characterization of generic independence as well as rigidity.

In a previous paper \cite{sitharam2014incidence},
we used pinned subspace-incidence systems towards solving the \emph{fitted
dictionary learning} problem,
i.e.\ dictionary learning with specified underlying hypergraph,
and gave a combinatorial characterization of minimal rigidity for a
more restricted version of pinned subspace-incidence system,
with $H$ being a uniform hypergraph and pins in $X$ being 1-dimension
subspaces. Moreover in a recent paper \cite{Baker2015}, the special case
of pinned line incidence systems was used to model biomaterials such as
cellulose and collagen fibrils in cell walls.
In this paper,
we extend the combinatorial characterization to general pinned
subspace-incidence systems,
with $H$ being a non-uniform hypergraph and pins in $X$ being subspaces
with arbitrary dimension.
As there are generally many data points per subspace in a dictionary
learning problem,
which can only be modeled with pins of dimension larger than $1$,
such an extension enables application to a much larger class
of fitted dictionary learning problems.

%We use combinatorial rigidity techniques to obtain a pure combinatorial characterization of pinned subspace-incidence systems 
%that are guaranteed to generically yield a locally unique realization,
%i.e.\ the systems that are generically minimally rigid. 

\end{abstract}

\section{Introduction}

A \emph{pinned subspace-incidence system} $(H,X)$ is an incidence constraint system
specified as a hypergraph $H$ together with a set $X$ of subspaces or \emph{pins} in $\R$
in one-to-one correspondence with the hyperedges of $H$.
A realization of $(H,X)$ assigns points in $\R$ to the vertices of $H$,
thereby subspaces to the hyperedges of $H$.
The subspace in $\R$ corresponding to a hyperedge of $H$ 
contains the associated pin from $X$. 
We are interested in characterization of
pinned subspace-incidence systems that are \emph{minimally rigid},
i.e.\ those systems that are guaranteed to generically yield a locally
unique realization.

 In a previous paper \cite{sitharam2014incidence}, 
%we gave a combinatorial characterization of minimal rigidity for a more restricted version of pinned subspace-incidence system, with $H$ being a uniform hypergraph and pins in $X$ being 1-dimension subspaces. 
we used pinned subspace-incidence systems towards solving \emph{fitted
 dictionary learning} problems, i.e.\ dictionary learning with specified underlying hypergraphs.
%Pinned subspace-incidence systems first  arise as the underlying structure of  
%\emph{fitted dictionary learning} problems \cite{sitharam2014incidence}, i.e.\ dictionary learning with specified underlying hypergraph.
\emph{Dictionary learning} (aka sparse coding) is the problem of obtaining a set \emph{dictionary vectors} that sparsely represent a set of given data points in $\R$. 
Geometrically, such a sparse representation can be viewed as a subspace arrangement  spanned by the dictionary vectors that contains all the data points. 
In  fitted dictionary learning,
 the underlying hypergraph $H$ of the subspace arrangement is specified, 
 and the problem becomes a pinned subspace-incidence systems
 with the pins corresponding to the span of data points on each subspace.

 Moreover in a recent paper \cite{Baker2015}, 
we have used pinned subspace-incidence systems in modeling biomaterials
such as cross-linking  cellulose and collagen microfibrils in cell walls~\cite{buehler2008nanomechanics,fall2013physical,smith1971plant}.
In such materials, each fibril is attached to some fixed larger organelle/membrane at one site,
and cross-linked at two locations with other fibrils.
Consequently, they can be modeled using a %simplified 
pinned line-incidence system with $H$ being a graph, %and pins in $X$ being 1-dimensional subspaces, 
where each fibril is modeled as an edge of $H$ with the two cross-linkings as its two vertices,
and the attachment is modeled as the corresponding pin.

We gave in \cite{sitharam2014incidence} a combinatorial characterization of minimal rigidity for a
  restricted version of pinned subspace-incidence system,
  with the underlying hypergraph $H$ being a uniform hypergraph and pins in $X$ being 1-dimension subspaces.

\section{Contributions}

In this paper, we extend the combinatorial characterization of minimal rigidity
to general pinned subspace-incidence systems, 
where $H$ can be any non-uniform hypergraph, and each pin in $X$ is a subspace with arbitrary dimension.
%with the $H$ being a non-uniform hypergraph and each pin being a subspace with arbitrary dimension.
Such an extension enables application to a much larger class of fitted dictionary learning problems,
since there are generally many data points per subspace in a dictionary
learning problem, which can only be modeled with pins of dimension larger than $1$.

As in our previous paper  \cite{sitharam2014incidence},
we apply the classic method of White and Whiteley \cite{white1987algebraic}
to combinatorially characterize the rigidity of general pinned subspace-incidence systems.
The primary technique is using the Laplace decomposition of the \emph{rigidity matrix}, 
which corresponds to a \emph{map-decomposition}~\cite{streinu2009sparse} of the underlying hypergraph. 
%to combinatorially capture the generic rigidity of the system. 
The  polynomial resulting from the Laplace decomposition is called the
\emph{pure condition}, which characterizes the conditions that the framework has 
to avoid for the combinatorial characterization to hold.
%and give so-called pure conditions that capture \emph{non-genericity}. 
%Note that the geometric conditions corresponding to the pure condition are not always clear,
%as shown by examples given in Section \ref{sec:proof}. 

Previous works on related types of geometric constraint frameworks
include pin-collinear body-pin frameworks~\cite{jackson2008pin}, 
direction networks~\cite{whiteley1996some},
slider-pinning rigidity~\cite{streinu2010slider},
body-cad constraint system~\cite{haller2012body},
$k$-frames~\cite{white1987algebraic,white1983algebraic}, 
and affine rigidity~\cite{gortler2013affine}.
However, we are not aware of any previous results on systems that are similar to pinned subspace-incidence systems.

%Pinned subspace-incidence frameworks are generalizations of related types of frameworks, 
%such as in pin-collinear body-pin frameworks~\cite{jackson2008pin}, 
%direction networks~\cite{whiteley1996some},
%slider-pinning rigidity~\cite{streinu2010slider},
%the molecular conjecture in 2D~\cite{servatius2006molecular},
%body-cad constraint system~\cite{haller2012body},
%$k$-frames~\cite{white1987algebraic,white1983algebraic}, 
%and affine rigidity~\cite{gortler2013affine}.

\section{Preliminaries}

In this section, we introduce the formal definition of pinned subspace-incidence systems
and basic concepts in combinatorial rigidity. 
% $\proj$.

A \emph{hypergraph} $H=(V,E)$ is a set $V$ of vertices and a set $E$ of hyperedges, 
where each hyperedge is a subset of $V$. 
The \emph{rank} $r(H)$ of a hypergraph $H$ is the maximum cardinality of any edge in $E$,
i.e.\ $r(H) = \max_{e_k \in E} s(e_k)$, where $s(e_k)$ denotes the cardinality of the hyperedge $e_k$.
A hypergraph is \emph{$s$-uniform} if all edges in $E$ have the same cardinality $s$.
A \emph{configuration} or \emph{realization} of a hypergraph $H = (V,E)$ in  $\R$
is a mapping of points in $\R$ to the vertices of $H$,  $p: V \rightarrow \R$. 
When there is no ambiguity, we simply use $p_i$ to denote the point $p(v_i)$, 
 $p(e_k)$ to denote the set of points $\{p(v_i) | v_i \in e_k\}$,
and $s_k$ to denote the cardinality $s(e_k)$.
%We  use $s(e_k)$ or simply $s_k$ to denote the cardinality of a hyperedge $e_k$ of $H$.

\begin{figure}

\centering
\begin{subfigure}{.4\linewidth}
  \centering
  \includegraphics[width=.8\linewidth]{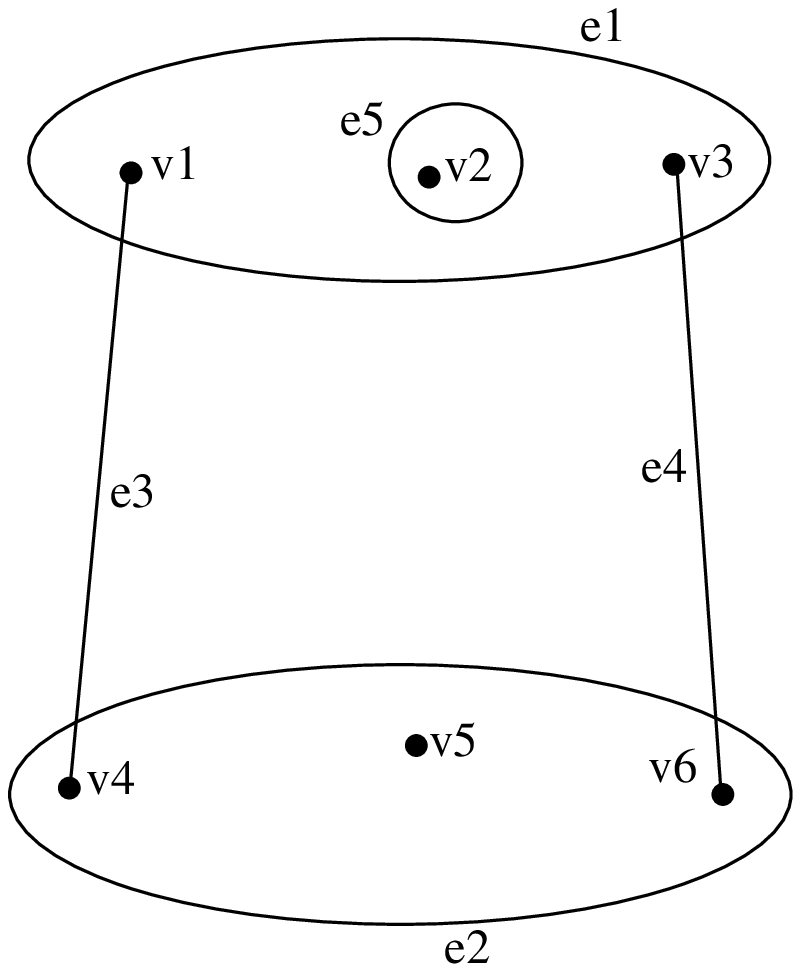}
  \caption{}
  \label{fig:hypergraph}
\end{subfigure}%
\begin{subfigure}{.4\linewidth}
  \centering
  \includegraphics[width=.8\linewidth]{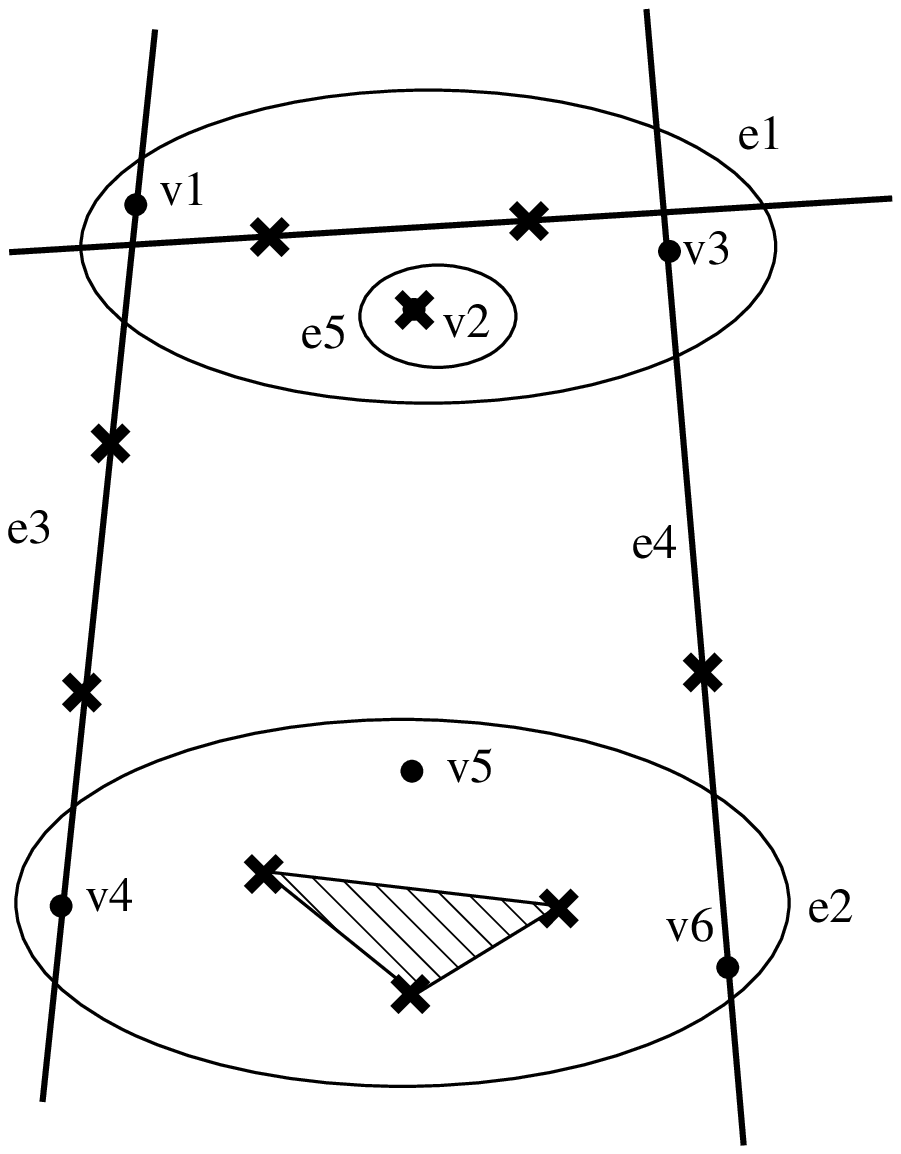}
  \caption{}
  \label{fig:pinned_hypergraph}
\end{subfigure}
\caption{(a) A rank-3 non-uniform hypergraph with 6 vertices and 5 edges, where $s_1 = s_2 = 3$, $s_3 = s_4 = 2$, $s_5 = 1$. 
(b) A pinned subspace-incidence framework in $d=4$ (projectivized in $\proj[3]$) on the hypergraph from (a), with $m_1 = m_3 = 2$,   $m_2 = 3$, $m_4 = m_5 = 1$, where the crosses on each hyperedge represent points spanning the associated pin.
}
\end{figure}

An example of a rank-3 hypergraph is given in Figure~\ref{fig:hypergraph}.

%A set $S$ of $s$ points in $\mathbb{P}^{d-1}(\mathbb{R})$ spans 
%a projective subspace of dimension $s-1$, which corresponds to a vector subspace of dimension $s$ in $\mathbb{R}^d$. 
In the following,  we use $\langle S \rangle$ to denote the subspace \emph{spanned} by  a set $S$ of points in $\R$.
%Note this subspace has dimension $\dim \langle S \rangle \le \min(|S|, d)$.

\begin{definition} [Pinned Subspace-Incidence System]
%Let $X$ be a given set of $m$ points in $\proj$, called a set of \emph{pins}.
A \emph{pinned subspace-incidence system} in  $\R$ is a pair $(H, X)$, 
where $H=(V,E,m)$ is a weighted hypergraph with hyperedges of rank  $r(H) < d$,
and $X = \{x_1, x_2, \ldots, x_{|E|}\}$ is a set of \emph{pins} (subspaces of $\R$)
in one-to-one correspondence with the hyperedges of $H$.
Here the weight assignment is a function $m: E(H) \rightarrow \mathbb{Z}^+$, 
where  $m(e)$ denotes the dimension of the  pin associated with the hyperedge $e$. 
Often we ignore the weight $m$ and just refer to the hypergraph $(V,E)$ as $H$.
%with the constraint that $x_i$ is contained in  $\langle p(e_i) \rangle$, the subspace  spanned by vertices of the corresponding hyperedge $e_i$.

A \emph{pinned subspace-incidence framework realizing} the pinned subspace-incidence system  $(H,X)$ is a triple $\framework$, 
where $p$ is a realization of $H$,
%The framework $\framework$ is a  \emph{realization} or \emph{solution} of $(H,X)$ if 
such that for all pins $x_k \in X$, 
$x_k$  is contained in $\langle p(e_k) \rangle$, the subspace  spanned by the set of points realizing the vertices of the  hyperedge $e_i$ corresponding to $x_k$.
%A \emph{pinned subspace-incidence framework} is a triple $\framework$, 
%where $p$ is a realization of $H$.
%The framework $\framework$ is a  \emph{realization} or \emph{solution} of $(H,X)$ if for all pins $x_i \in X$, 
%$x_i$  is contained in $\langle p(e_i) \rangle$, the subspace  spanned by the set of points realizing the vertices of the  hyperedge $e_i$ corresponding to $x_i$.
\end{definition}

We may write $m(x_k)$ or simply $m_k$ in substitute of $m(e_k)$, where $x_k$ is the pin associated with the hyperedge $e_k$.

Since we only care about incidence relations, 
we projectivize the Euclidean space $\R$ to treat the pinned subspace-incidence system in the real projective space $\proj$,
%(i.e.\ the projectivization of the Euclidean space $\R$),
and use the same notation for the pins and hypergraph realization when the meaning is clear from the context.

Figure~\ref{fig:pinned_hypergraph} gives an example of a pinned subspace-incidence framework in the projective space $\proj$ with $d=4$, where each pin is the subspace spanned by the set of cross-denoted points on the corresponding hyperedge.

%\begin{definition}
%Given a pinned subspace-incidence system $(H, X)$,
%the \emph{underlying pinned subspace-incidence graph} is the pair $(H, m)$,
%where $m: E(H) \rightarrow \mathbb{Z}^+$ is a function mapping each hyperedge $e_k$ of $H$ 
%to the dimension $m_k$ of the associated pin in $X$. 
%% $H=(V,E)$ of rank $s$ together with a function $\mathcal{P}: E \rightarrow \mathbb{Z^+}$ specifying a positive integer for each hyperedge,  such that $\mathcal{P}(e) \le |e|$ for every $e \in E$.
%\end{definition}
%
%\begin{definition}
%Two frameworks $(H_1,X_1,p_1)$ and $(H_2,X_2,p_2)$ are \textit{equivalent} if
%$H_1=H_2$ and $X_1 =X_2$, i.e.\ they are realizations for the same pinned subspace-incidence system.
%%satisfy the same algebraic equations for the same labeled hypergraph and ordered set of pins. 
%They are \emph{congruent} if they are equivalent and $p_1 = p_2$.
%\end{definition}

\begin{definition}
A pinned subspace-incidence system $(H,X)$ is \emph{independent} if none of the algebraic constraints is in the ideal generated by others, which generically implies the existence of a realization.
It is \emph{rigid} if there exist at most finitely many realizations. 
It is \emph{minimally rigid} if it is both rigid and independent. 
It is \emph{globally rigid} if there exits at most one realization. 
%
%A pinned subspace-incidence framework $\framework$ is \emph{rigid} (i.e.\ \emph{locally unique}) if there is a neighborhood $N(p)$ of $p$, 
%such that any framework $\framework[q]$ equivalent to $\framework$ with $q \in N(p)$ is also congruent to $\framework$. 
%A rigid framework $\framework$ is \emph{minimally rigid} if it becomes flexible after removing any pin.
%A framework $\framework$ is \emph{globally rigid} (i.e.\ \emph{globally unique}) if any framework equivalent to $\framework$ is also congruent to $\framework$. 
\end{definition}

\section{Algebraic Representation and Linearization}

In the following,  
we use $A[R,C]$ to denote a submatrix of a matrix $A$, 
where $R$ and $C$ are respectively index sets of the rows and columns contained in the submatrix.
In addition, $A[R,\Cdot]$ represents the submatrix containing row set $R$ and all columns, 
and $A[\Cdot,C]$ represents the submatrix containing column set $C$ and all rows.

\subsection{Algebraic Representation}

A pin $x_k$ associated with a hyperedge  $e_k = \{v^k_1, v^k_2, \ldots, v^k_{s_k}\}$ of cardinality $s_k$ 
is constrained to be contained in the subspace $\langle p(e_k) \rangle$ spanned by the point set  $ \{p^k_1, p^k_2, \ldots, p^k_{s_k}\}$. 
%with the homogeneous coordinates $x = (x_1, x_2, \ldots, x_{d-1}, 1)$, $v_i = (v_{i,1}, v_{i,2}, \ldots, v_{i,{d-1}},1)$.
%
As $x_k$ is a subspace of dimension $m_k - 1$ in $\proj$,
we can pick a set of $m_k$ points $\{x^k_1, x^k_2, \ldots, x^k_{m_k}\}$ spanning $x_k$ from $x_k$.
Now the constraint is equivalent to requiring each such point $x^k_l$ to lie on $\langle p(e) \rangle$, for  $1 \le l \le m_k$.
% $x_1, x_2, \ldots x_{m(x)}$ all lie on  
We call each such point $x^k_l$ a \emph{multipin}
as it acts like a pin with  $m(x^k_l) = 1$.

\sloppy Using homogeneous coordinates
$p^k_i = [\begin{array}{cccc}p^k_{i,1} & p^k_{i,2} & \ldots & p^k_{i,d-1}\end{array}] $ and
$x^k_l = [ \begin{array}{cccc}x^k_{l,1} & x^k_{l,2} & \ldots & x^k_{l,d-1} \end{array}]$,
we write this incidence constraint for each point $x_l$ 
by letting 
all the $s_k \times s_k$ minors of the $s_k \times (d-1) $ matrix
\[
E^k_l=
\left[
\begin{array}{c}
p^k_1 - x^k_l	\\
 p^k_2 -x^k_l \\
\vdots \\
 p^k_{s_k} - x^k_l 
\end{array} \right]
\]
\fussy
be zero. There are $d-1 \choose s_k$ minors, giving $d-1 \choose s_k$ equations. 
%\begin{equation} \label{eq:constraint}
%\det\left(E[\Cdot, C(t)]\right) = 0, \qquad 1 \le t \le {d-1 \choose s}
%\end{equation} 
%where $C(t)$ enumerates all the $s$-subsets of columns of $E$. 
%%The set $R(t)$ contains indices  $\{l_1,l_2, \ldots, l_s \}$.
%
Note that any $d-s_k$ of these $d-1 \choose s_k$ equations are independent and span the rest.
%TODO why
%since $\langle p(e) \rangle$
%is a $s$-dimensional subspace in a $d$-dimensional space,
%which only has $s(d-s)$ degrees of freedom.
%TODO maybe need to rearrange here
So we can write the incidence constraint as $(d-s_k)$ independent equations: 
\begin{equation} \label{eq:constraint}
\det\left(E^k_l[\Cdot, C(t)]\right) = 0, \qquad 1 \le t \le d-s_k
\end{equation} 
where $C(t)$ denote the following index sets of columns in $E$:
\[
C(t) = \{1, 2, \ldots s_k-1\} \cup \{ s_k-1+t \} %= \mathbf{S} \cup \{s-1+t\},
\qquad 1 \le t \le d-s
\]
%where $\mathbf{S}$ denotes the set $\{1, 2, \ldots s-1\}$.
In other words, $C(t)$ contains the first $s_k-1$ columns together with Column $s_k-1+t$.

Now
the incidence constraint for the pin $x_k$ is represented
as $m_k (d-s_k)$ equations
for all the $m_k$ multipins $\{x^k_1, x^k_2, \ldots x^k_{m_k}\}$.
Consequently, the pinned subspace-incidence  problem reduces to solving a system 
of  $\sum_{k=1}^{|E|} m_k (d-s_k)$ equations, each of form~\eqref{eq:constraint}.  
We denote this algebraic system by $\algesystem = 0$.

\subsection{Genericity}

We are interested in characterizing \emph{minimal rigidity} of pinned subspace-incidence systems.
However, checking independence relative to the ideal generated by the variety is computationally hard and best known algorithms, such as computing Gr\"{o}bner basis, are exponential in time and space \cite{mittmann2007grobner}.
However, the algebraic system can be linearized at \emph{generic} or \emph{regular} (non-singular) points,
whereby the  independence and rigidity of the algebraic system $\algesystem = 0$ 
reduces to linear independence and maximal rank at \emph{generic} frameworks. 

In algebraic geometry, a property being generic intuitively means that the property holds on the open dense complement of an (real) algebraic variety. Formally, 

\begin{definition}
\label{def:generic_framework}
A pinned subspace-incidence system $(H,X)$ is \emph{generic} w.r.t.\ a property $Q$ if and only if there exists a neighborhood $N(X)$ such that for all systems $(H,X')$ with $X' \in N(X)$,
$(H,X')$ satisfies $Q$ if and only if $(H,X)$ satisfies $Q$.

Similarly, a framework $\framework$ is \emph{generic} w.r.t.\ a property $Q$ if and only if there exists a neighborhood $N(p)$ such that for all frameworks $\framework[q]$ with $q \in N(p)$,
$\framework[q]$ satisfies $Q$ if and only if $\framework$ satisfies $Q$.
\end{definition}

%Furthermore we can define generic properties in terms of the combinatorics of the pinned subspace-incidence system, 
%i.e.\ the underlying hypergraph $H$ and the dimension of the pins associated with the hyperedges of $H$.
Furthermore we can define generic properties in terms of the underlying weighted hypergraph.

\begin{definition}
\label{def:generic_property}
A property $Q$ of pinned subspace-incidence systems is \emph{generic} (i.e, becomes a property of the underlying weighted hypergraph alone) if for any weighted hypergraph $H = (V,E,m)$, 
%pinned subspace-incidence systems $(H,X)$
%with the same hypergraph $H$ and the same dimension for each hyperedge's associated pin,
% are the same and associated pins for each hyperedge have the same dimension,
 either all generic (w.r.t.\ $Q$) systems $(H,X)$ satisfies $Q$, or all generic (w.r.t.\ $Q$) systems $(H,X)$ do not satisfy $Q$.
\end{definition}

%A framework $\framework$ is \emph{generic} for property $Q$ if  an algebraic variety $V_Q$ specific to $Q$ is
%avoided by the given framework $\framework$.
%Often, for convenience in relating $Q$ to other properties, a more
%restrictive notion of genericity is used than stipulated by Definition \ref{def:generic_framework} or \ref{def:generic_property},
%i.e.\ another variety $\acute{V}_Q$ is  chosen so that
%$V_Q \subseteq \acute{V}_Q$, as in Lemma \ref{lem:generic}. 
%Ideally, the variety $\acute{V}_Q$ corresponding to the chosen notion of genericity
%should be as tight as possible for the property Q (necessary and
%sufficient for Definition \ref{def:generic_framework} and \ref{def:generic_property}), 
%and should be explicitly defined, or at
%least easily testable for a given framework.

%TODO do we need this?
Once an appropriate notion of genericity is defined, 
we can treat $Q$  as a property of a hypergraph.
The primary activity of the area of combinatorial rigidity is to
  give purely combinatorial characterizations of such generic
properties $Q$.
In the process of drawing such combinatorial characterizations,
the notion of genericity  may have to be further restricted %, i.e.\ the variety $\acute{V}_Q$ is further expanded 
by so-called pure conditions that are necessary for the combinatorial characterization to go through 
(we will see this in the proof of Theorem \ref{thm:rigidity_condition}).

\subsection{Linearization as Rigidity Matrix}

Next we follow the approach taken by  traditional combinatorial rigidity theory \cite{asimow1978rigidity,graver1993combinatorial} 
to show that rigidity and independence (based on nonlinear polynomials) of pinned subspace-incidence systems 
are generically properties of the underlying weighted hypergraph $H$,
and can furthermore be captured by linear conditions in an infinitesimal setting.
Specifically, Lemma~\ref{lem:generic} shows that
rigidity of a pinned subspace-incidence system
is equivalent to existence of a full rank  \emph{rigidity matrix}, %of a generic framework,
obtained by taking the Jacobian of the algebraic system $\algesystem$ at a regular point.

A \emph{rigidity matrix} of a framework $\framework$ is a matrix  whose kernel is the infinitesimal motions (flexes) of $\framework$. 
%A framework is \emph{infinitesimally independent} if the rows of the rigidity matrix are independent.
A framework is \emph{infinitesimally rigid} if the space of infinitesimal motion is trivial, 
i.e.\ the rigidity matrix has full rank. 
%A framework is \emph{infinitesimally minimally rigid} if it is both infinitesimally independent and rigid.
%
To define a rigidity matrix for a pinned subspace-incidence framework $\framework$,
we take the Jacobian of the algebraic system $\algesystem = 0$
by taking partial derivatives with respect to the coordinates of $p_i$'s. 
In the Jacobian, each vertex $v_i$ has $d-1$  corresponding columns, 
and %each pin $x_k$ has %$d-1 \choose s$ corresponding equations, thus 
%$m_k(d-s_k)$ corresponding rows.
each pin $x_k$ associated with the hyperedge  $e_k = \{v^k_1, v^k_2, \ldots, v^k_{s_k} \}$
has $m_k(d-s_k)$ corresponding rows, 
where each equation $\det\left(E^k_l[\Cdot, C(t)]\right)=0$ \eqref{eq:constraint},
i.e.\ Equation $t$ of the multipin $x^k_l$,
gives the following row
(the columns corresponding to vertices not in $e_k$ are all zero): %in the Jacobian
%(where  $x_k$ lies on the subspace spanned by  $e_k = \{v^k_1, v^k_2, \ldots, v^k_s \}$):

\fussy
\begin{align}
\label{eq:row}
\bigg[\; 0,&\ldots,0,0, \frac{\partial \det\left(E^k_l[\Cdot, C(t)]\right)}{\partial p^k_{1,1}}, \frac{\partial \det\left(E^k_l[\Cdot, C(t)]\right)}{\partial p^k_{1,2}}, \ldots,\frac{\partial \det\left(E^k_l[\Cdot, C(t)]\right)}{\partial p^k_{1,d-1}}, 0,0, \notag \\ 
&\ldots,0,0,\frac{\partial \det\left(E^k_l[\Cdot, C(t)]\right)}{\partial p^k_{2,1}}, \frac{\partial \det\left(E^k_l[\Cdot, C(t)]\right)}{\partial p^k_{2,2}},  \ldots, \frac{\partial \det\left(E^k_l[\Cdot, C(t)]\right)}{\partial p^k_{2,d-1}}, 0,0,\ldots \notag \\ 
&\ldots \ldots \notag \\ 
&\ldots, 0,0, \frac{\partial \det\left(E^k_l[\Cdot, C(t)]\right)}{\partial p^k_{s_k,1}}, \frac{\partial \det\left(E^k_l[\Cdot, C(t)]\right)}{\partial p^k_{s_k,2}},  \ldots, \frac{\partial \det\left(E^k_l[\Cdot, C(t)]\right)}{\partial p^k_{s_k,d-1}}, 0,\ldots,0 \; \bigg]
\end{align}

Let $V^k$ be the matrix whose rows are coordinates of $p^k_1, p^k_2, \ldots, p^k_{s_k}$:
\[
\left[
\begin{array}{cccc}
p_{1,1} & p_{1,2} & \ldots & p_{1, d-1}\\
p_{2,1} & p_{2,2} & \ldots & p_{2, d-1}\\
\vdots & \vdots & \ddots & \vdots \\
p_{s_k,1} & p_{s_k,2} & \ldots & p_{s_k, d-1}
\end{array}
\right]
\]
Let $V^k_t$ be the $V^k[\Cdot, C(t)]$, i.e.\ the $s_k \times s_k$ submatrix of $V^k$ containing only columns in $C(t)$.
Let $V^k_{t,j}$ be the matrix obtained from $V^k_t$ by replacing the column corresponding to Coordinate $j$ with the all-ones vector $(1, 1, \ldots, 1)$ for $j \in C(t)$, and the zero matrix for $j \notin C(t)$. 
Let $D^k_{t,j}$ be the determinant of $V^k_{t,j}$.
%Starting from a frame realizing $(H,X)$ xxxx ...
Let $x_l  = \sum_{i=1}^{s_k} b^{k,l}_i v^k_i$
(note that $\sum_{i=1}^{s_k} b^{k,l}_i = 1$).
%
%Let $V$ be the volume of simplex formed by vertices in $e_k$.
%For $1 \le j \le d-1$, we define
%\[
%V^k_{t,j} = 
%\begin{cases}
%V\text{ projected on coordinate set } C(t) \setminus \{j\} & j \text{ in } C(t)\\
% 0 & j \notin C(t)
%\end{cases}
%\]
%
% NOTE: for the same k:
% - D is related to t and j, not to l
% - b is related to l and i?, not to t
Now \eqref{eq:row} can be rewritten in the following simplified form:

%\begin{equation}
%\hspace{-30pt}
%\scalemath{0.9}{
%\begin{blockarray}{*{16}{c}}
% & v^k_{1,1} & v^k_{1,2} & \ldots & v^k_{1,d-1} & \ldots\ldots
% & v^k_{1,1} & v^k_{1,2} & \ldots & v^k_{1,d-1} & \ldots\ldots\ldots
% & v^k_{s_k,1} & v^k_{s_k,2} & \ldots & v^k_{s_k,d-1} &   \\
%\begin{block}{[*{16}{c}]}
%\ldots & D^k_{t,1} b^{k,l}_1 & D^k_{t,2} b^{k,l}_1 & \ldots & D^k_{t,d-1} b^{k,l}_1 & \ldots\ldots
%    & D^k_{t,1} b^{k,l}_2 & D^k_{t,2} b^{k,l}_2 & \ldots & D^k_{t,d-1} b^{k,l}_2 & \ldots\ldots
%     \ldots & D^k_{t,1} b^{k,l}_{s_k} & D^k_{t,2} b^{k,l}_{s_k} & \ldots & D^k_{t,d-1} b^{k,l}_{s_k} &\ldots  \\
%\end{block}
%\end{blockarray}
%}
%\end{equation}

\begin{align}
r^k_{t,l} = \Big[\;0,\, &\ldots,\, 0,\, 0,\,   D^k_{t,1} b^{k,l}_1,\,  D^k_{t,2} b^{k,l}_1,\,  \ldots,\,  D^k_{t,d-1} b^{k,l}_1,\,  0,\, 0,\, \notag\\
&\ldots,\, 0,\, 0,\,  D^k_{t,1} b^{k,l}_2,\,  D^k_{t,2} b^{k,l}_2,\,  \ldots,\,  D^k_{t,d-1} b^{k,l}_2,\,  0,\, 0,\, \ldots \notag\\
&\ldots \ldots \notag\\
&\ldots,\,  0,\, 0,\,  D^k_{t,1} b^{k,l}_{s_k},\,  D^k_{t,2} b^{k,l}_{s_k},\,  \ldots,\,  D^k_{t,d-1} b^{k,l}_{s_k},\,  0,\, 0,\, \ldots,\, 0 \;\Big] \label{eq:row_pattern}
\end{align}
%where $\sum_{1 \le i \le s} b^k_i = 1$.

Each vertex $v^k_i$ has the entries $D^k_{t,j} b^{k,l}_i, 1 \le j \le d-1$ in its $d-1$ columns, among which exactly $s_k$ entries with $j \in C(t)$, i.e.\ the first $s_k-1$ columns together with Column $s_k-1+t$, are generically non-zero.
%
%Each vertex $v^k_{i}$ has the entries 
%$(-1)^q V^k_{t, j} b^k_i$ in its $1 \le j \le d-1$ columns, among which $s$ entries (where $j \in C(t)$, i.e.\ the first $s-1$ columns together with Column $s-1+t$) are generically non-zero.
%Here $q = j+1$ for $j \le s-1$, $q = d-s+1$ for $j \ge s$.
%
Note that %for $j \ge s$, the only non-zero column is $j = s_k-1+t$ with the entry $D^k_{t, s_k-1+t}$, 
%As $C(t) \setminus j = \{1, 2, \ldots, s_k-1\}$ for all $t$.
%and from the definition above we can see that
the terms $D^k_{t, s_k-1+t}$  are equal for all $t$, 
so we may just use $D^k$ to denote it.

For each $1 \le t \le d-s_k$, 
there are $m_k$ rows  as \eqref{eq:row_pattern}, where each multipin $x^k_l$ corresponds to the row $r^k_{t, l}$ for $1 \le l \le m_k$.
These $m_k$ rows have exactly the same row pattern 
except for different $b^{k,l}_i$'s:
%When there are $m_k \le s$ pins on the same hyperedge, we have the following row form
%for each $1 \le t \le d-s$:
\[
\hspace{-50pt}
\begin{blockarray}{*{12}{c}}
 & v_{1,1} & v_{1,2} & \ldots & v_{1,d-1} & & & v_{s_k,1} & v_{s_k,2} & \ldots & v_{s_k,d-1} & \\
 \begin{block}{[>{\hspace{0.5em}}*{12}{c}<{\hspace{0.5em}}]}
\ldots & D^k_{t,1} b^{k,1}_1 & D^k_{t,2} b^{k,1}_1 & \ldots & D^k_{t,d-1} b^{k,1}_1 &
     \ldots & 
     \ldots & D^k_{t,1} b^{k,1}_{s_k} & D^k_{t,2} b^{k,1}_{s_k} & \ldots & D^k_{t,d-1} b^{k,1}_{s_k} &\ldots \\
          \ldots & D^k_{t,1} b^{k,2}_1 & D^k_{t,2} b^{k,2}_1 & \ldots & D^k_{t,d-1} b^{k,2}_1 &
               \ldots & 
               \ldots & D^k_{t,1} b^{k,2}_{s_k} & D^k_{t,2} b^{k,2}_{s_k} & \ldots & D^k_{t,d-1} b^{k,2}_{s_k} &\ldots \\
     \BAmulticolumn{12}{c}{\ddots}\\
     \ldots & D^k_{t,1} b^{k,m_k}_1 & D^k_{t,2} b^{k,m_k}_1 & \ldots & D^k_{t,d-1} b^{k,m_k}_1 &
          \ldots & 
          \ldots & D^k_{t,1} b^{k,m_k}_{s_k} & D^k_{t,2} b^{k,m_k}_{s_k} & \ldots & D^k_{t,d-1} b^{k,m_k}_{s_k} &\ldots \\
 \end{block}
\end{blockarray}
\]

\begin{example}
For $d=4$, consider a pin $x$ with $m(x) = 2$ associated with the hyperedge $e = \{v_1, v_2\}$.
The pin has the following $m(x) \cdot (d - s(e)) = 4$ rows in the simplified Jacobian (the index $k$ is omitted):
\[
\begin{blockarray}{c@{\hspace{1.5em}}c@{\hspace{1.5em}}c@{\hspace{1.5em}}c@{\hspace{1.5em}}c@{\hspace{1.5em}}c@{\hspace{1.5em}}c@{\hspace{1.5em}}c@{\hspace{1.5em}}c@{\hspace{1.5em}}c@{\hspace{1.5em}}}
& & v_{1,1} & v_{1,2} & v_{1,3} & & v_{2,1} & v_{2,2} & v_{2,3} & \\
\begin{block}{c@{\hspace{1.5em}}[c@{\hspace{1.5em}}c@{\hspace{1.5em}}c@{\hspace{1.5em}}c@{\hspace{1.5em}}c@{\hspace{1.5em}}c@{\hspace{1.5em}}c@{\hspace{1.5em}}c@{\hspace{1.5em}}c@{\hspace{1.5em}}]}
t=1, l=1 & \quad\ldots & D_{1,1} b^1_1& D b^1_1& 0 & \ldots & D_{1,1} b^1_2 & D b^1_2 & 0 & \ldots\quad \\
t=1, l=2 & \quad\ldots & D_{1,1} b^2_1 & D b^2_1 & 0 & \ldots & D_{1,1} b^2_2 & D b^2_2 & 0 & \ldots\quad \\
t=2, l=1 & \quad\ldots & D_{2,1} b^1_1& 0 & D b^1_1& \ldots & D_{2,1} b^1_2 & 0 & D b^1_2 & \ldots\quad \\
t=2, l=2 & \quad\ldots & D_{2,1} b^2_1 & 0 & D b^2_1 & \ldots & D_{2,1} b^2_2 & 0 & D b^2_2 & \ldots\quad \\
\end{block}
\end{blockarray} 
\]
\end{example}

We define the \emph{rigidity matrix} $M\framework$ or simply $M(p)$ for a pinned subspace-incidence framework $\framework$ to be the simplified Jacobian matrix obtained above, where each row has form \eqref{eq:row_pattern}.
It is a matrix of size $\sum_k m_k (d - s_k)$ by $n(d-1)$.

\begin{definition} \label{def:genericity}
A pinned subspace-incidence framework $\framework$ and the corresponding system $(H,X)$ is \emph{generic} 
%todo{w.r.t. ?} 
if $p$ and $X$ are regular\slash{}non-singular points with respect to the algebraic system $\algesystem=0$.
\end{definition}
We use $M(H)$ of simply $M$ to denote the generic rigidity matrix for a weighted hypergraph $H$.
Note that the rank of $M$ cannot be less than the rank of $M(p)$ for any specific realization $p$. 
%todo{is this implied from the definition? genericity w.r.t. inf rigid}

\begin{lemma}
\label{lem:generic}
%If $p$ and $X$ are regular\slash{}non-singular with respect to the system $\algesystem = 0$,
%then 
Generic infinitesimal rigidity of a pinned subspace-incidence framework $\framework$ is equivalent to generic rigidity of the system $(H,X)$.
%\todo{weighted hypergraph $H$? system $(H,X)$?}.
\end{lemma}

The proof of Lemma \ref{lem:generic} follows the traditional combinatorial rigidity approach~\cite{asimow1978rigidity} and is given in the Appendix. 

%todo{\begin{lemma}
%Rigidity is a generic property of the weighted hypergraph $H$.
%(One generic system rigid $\implies$ all generic systems rigid)
%\end{lemma}}

\section{Combinatorial rigidity characterization}

\subsection{Required hypergraph properties}

This section introduces pure hypergraph properties and definitions that will be used in stating and proving our main theorem.

\begin{definition} \label{def:tightness}
A hypergraph $H=(V,E)$ is $(k,0)$-sparse if for any ${V'} \subset V$, 
the induced subgraph ${H'}=({V'},{E'})$ satisfies $|{E'}| \leq k|{V'}|$. 
A hypergraph $H$ is $(k,0)$-tight if $H$ is $(k,0)$-sparse and $|E| = k|V|$.
 %$\acute{G}$ is the vertex induced subgraph augmented with a self-loop at $v_i$ 
%       when there are two edges between the same vertex $v_j \in V - \acute{V}$ to the same vertex $v_i \in \acute{V}$.
\end{definition}

This is a special case of the $(k,l)$-sparsity condition 
that was formally studied widely in the geometric constraint solving and combinatorial rigidity literature 
before it was given a name in \cite{lee2007graded}.
A relevant concept from graph matroids is \emph{map-graph}, defined as follows.

\begin{definition}
An \emph{orientation} of a hypergraph is given by identifying as the \emph{tail} of each edge one of its endpoints.
The \emph{out-degree} of a vertex is the number of edges which identify it as the tail and connect $v$ to $V - v$.
A \emph{map-graph} is a hypergraph that admits an orientation such that the out degree of every vertex is
exactly one. 
\end{definition}

The following lemma from \cite{streinu2009sparse} follows Tutte-Nash Williams \cite{tutte1961problem,nash1961edge}
to give a useful characterization of $(k,0)$-tight graphs in terms of maps.

\begin{lemma} 
\label{lem:map_decomposition}
A hypergraph $H$ is composed of $k$ edge-disjoint map-graphs if and only if $H$ is $(k,0)$-tight.
\end{lemma}

Our characterization of rigidity of a weighted hypergraph $H$ is based on map-decomposition of
a \emph{multi-hypergraph} $\hat{H}$ obtained from $H$.

\begin{definition}
Given a weighted hypergraph $H = (V, E, m)$,
the associated \emph{multi-hypergraph} $\hat{H} = (V, \hat{E})$ is obtained by 
replacing each hyperedge $e_k$ in $E$ with a set ${E}^k$ of $m_k (d-s_k)$ copies of \emph{multi-hyperedges}.
%%each hyperedge $e_k$ has $m_k$ multiples, and each multiple has $d-s_k$ copies.
%Each such set $E^k$ is divided into $m_k$ subsets ${E}^k_p, 1 \le p \le m_k$, each of size $d-s_k$, 
%and containing the edge copies ${e}^k_p(t), 1 \le t \le d-s_k$.
%%Each hyperedge $e_k$ corresponds to a subset $\hat{E}_k$ of $m_k$ hyperedges
%%$m_k$ multiples, and each multiple has $d-s_k$ copies.
\end{definition}

A \emph{labeling} of a multi-hypergraph $\hat{H}$ gives a one-to-one correspondence between $E^k$ and 
the set $R^k$ of $m_k (d-s_k)$ rows for the hyperedge $e^k$ in the rigidity matrix $M$,
where the multi-hyperedge corresponding to the row $r^k_{t,l}$ is labeled $e^k_{t,l}$.

%assigns each edge in $E^k$ two indices $l$ and $t$, $1 \le l \le m_k$, $1 \le t \le d-s_k$.
%A hyperedge $e^k_{t, l}$ in the labeled multi-hypergraph corresponds to the row with the corresponding $t$ and $l$ in the rigidity matrix.

\subsection{Characterizing rigidity}
\label{sec:proof}

In this section, we apply \cite{white1987algebraic} to give combinatorial characterization for minimal rigidity of pinned subspace-incidence systems.

\begin{theorem}[Main Theorem] \label{thm:rigidity_condition}
A pinned subspace-incidence system is generically minimally rigid
if and only if:
\begin{enumerate}

\item 
The underlying weighted hypergraph $H = (V,E, m)$ satisfies
 $\sum_{k=1}^{|E|} m_k(d-s_k) = (d-1)|V|$, 
and $\sum_{e_k \in E'} m_k(d-s_k) \le (d-1)|V'|$ for every vertex induced subgraph $H'=(V',E')$.
In other words, the associated multi-hypergraph $\hat{H} = (V, \hat{E})$ has a decomposition into $(d-1)$ maps. 

\item There exists a labeling of $\hat{H}$ \emph{compatible with the map-decomposition (defined later)}  such that in each set $E^k$ of multi-hyperedges,
%the map-decomposition has the following properties:

%Edge copies from the same subset ${E}^k_p$ 
%Duplicated edges from the same pin, i.e.\ rows with same $k$, same $p$, different $t$, 
(a) two multi-hyperedges $e^k_{t_1,l_1}$ and $e^k_{t_2,l_2}$ with $l_1 = l_2$
are not contained in the same map in the map-decomposition, 
%{There are at most $m_k$ multi-hyperedges from the same set $E^k$ that are included in the same map.}

(b) two multi-hyperedges $e^k_{t_1,l_1}$ and $e^{k}_{t_2,l_2}$ with $ t_1=t_2$ do not have the same vertex as tail in the map-decomposition. 
%The same duplicate from pins in the subspace, i.e.\ rows with same $k$, same $t$ and different $p$, don't have have the same vertex as tail.

%(c) at most $s_k -1$ maps can contain two multi-hyperedges $e^k_{t_1,l_1}$ and $e^{k}_{t_2,l_2}$ with $ t_1 \ne t_2$, $l_1 \ne l_2$.
\end{enumerate}

\end{theorem}

%% TODO should the pure condition be here?
%One thing to be noticed is the validity of Theorem~\ref{thm:rigidity_condition} requires the framework
%to avoid certain polynomial called \emph{pure condition} in addition of being generic as in Definition~\ref{def:genericity}
%(see Equation \eqref{eq:pure_condition}).
%The pure condition characterizes the badly behaved cases that break the combinatorial characterization of infinitesimal rigidity. 
%\todo{Geometric meaning not always explicit. }
%One particular condition not captured by Theorem~\ref{thm:rigidity_condition} but enforced by the pure condition is that 
%there cannot exists a subgraph $(V', E')$ of $H$ such that $\sum_{e_k \in E'} m_k > |V'|$,
%otherwise simple counterexamples can be constructed to the characterization of the main theorem: see Section~\ref{sec:open_problem}. 

To prove Theorem~\ref{thm:rigidity_condition}, we apply Laplace expansion to the determinant of the rigidity matrix $M$, 
which corresponds to decomposing the $(d-1,0)$-tight multi-hypergraph $\hat{H}$ as a union of $d-1$ maps.
We then prove $\det (M)$ is not identically zero by showing that the minors corresponding to each map are not identically zero, as long as a certain polynomial called \emph{pure condition} is avoided by the framework. 
The pure condition characterizes the non-genericity that the framework has to avoid in order for the combinatorial characterization to go through: see Example~\ref{ex:pure_cond}.

A Laplace expansion rewrites the determinant of the rigidity matrix $M$ 
as a sum of products of determinants (brackets) representing each of the coordinates taken separately. 
In order to see the relationship between the Laplace expansion and the map-decomposition, we first group the columns of $M$ into $d-1$ column groups $C_j$ according to the coordinates, 
where columns for the first coordinate of each vertex belong to $C_1$, 
columns for the second coordinate of each vertex belong to $C_2$, etc. 

%The $m_k$ rows corresponding to a specific hyperedge $e^k$ and a particular $t$ have the following form
%after regrouping:
%\[
% \def\arraystretch{1.5}
% \hspace{-20pt}
%\begin{blockarray}{*{12}{c}}
%  & \BAmulticolumn{4}{c}{C_1} &  &  & \BAmulticolumn{4}{c}{C_{d-1}} &\\[-4pt]
% & v_{1,1} & v_{2,1} & \ldots & v_{s_k,1} & & & v_{1,d-1} & v_{2,d-1} & \ldots & v_{s_k,d-1} & \\[1pt]
% \begin{block}{[c|cccc|cc|cccc|c]}
%\ldots & D^k_{t,1} b^{k,1}_1 & D^k_{t,1} b^{k,1}_2 & \ldots & D^k_{t,1} b^{k,1}_{s_k}  & \ldots & \ldots 
%& D^k_{t,d-1} b^{k,1}_1 & D^k_{t,d-1} b^{k,1}_2  & \ldots & D^k_{t,d-1} b^{k,1}_{s_k} & \ldots  \\
%\ldots & D^k_{t,1} b^{k,2}_1 & D^k_{t,1} b^{k,2}_2 & \ldots & D^k_{t,1} b^{k,2}_s  & \ldots & \ldots & D^k_{t,d-1} b^{k,2}_1 &
% D^k_{t,d-1} b^{k,2}_2  & \ldots & D^k_{t,d-1} b^{k,2}_s & \ldots  \\
% & &&&& \BAmulticolumn{2}{c|}{\ddots} &&&&&\\
%\ldots & D^k_{t,1} b^{k,m_k}_1 & D^k_{t,1} b^{k,m_k}_2 & \ldots & D^k_{t,1} b^{k,m_k}_s  & \ldots & \ldots & D^k_{t,d-1} b^{k,m_k}_1 &
% D^k_{t,d-1} b^{k,m_k}_2  & \ldots & D^k_{t,d-1} b^{k,m_k}_s & \ldots  \\
% \end{block}
%\end{blockarray} 
%\]

\begin{example} 
%TODO I think we need more than one hyperedge --- maybe a full graph
For $d=4$, consider a pin $x$ with $m(x)=2$ associated with the hyperedge $e = {v_1, v_2}$.
The regrouped rigidity matrix has $d-1 = 3$ column groups, where 
the pin $x$ has the following $4$ rows (the index $k$ is omitted):
\[
 \def\arraystretch{1.2}
\hspace{-30pt}
\begin{blockarray}{c@{\hspace{0.8em}}c@{\hspace{0.8em}}c@{\hspace{0.8em}}c@{\hspace{0.8em}}c@{\hspace{0.8em}}c@{\hspace{0.8em}}c@{\hspace{0.8em}}c@{\hspace{0.8em}}c@{\hspace{0.8em}}c@{\hspace{0.8em}}c@{\hspace{0.8em}}c@{\hspace{0.8em}}c@{\hspace{0.8em}}c@{\hspace{0.8em}}c@{\hspace{0.8em}}c@{\hspace{0.8em}}}
&  & v_{1,1} & & v_{2,1} & & & v_{1,2} && v_{2,2} & & & v_{1,3} && v_{2,3} & \\
\begin{block}{c@{\hspace{0.8em}}[c@{\hspace{0.8em}}c@{\hspace{0.8em}}c@{\hspace{0.8em}}c@{\hspace{0.8em}}c@{\hspace{0.5em}}|@{\hspace{0.5em}}c@{\hspace{0.8em}}c@{\hspace{0.8em}}c@{\hspace{0.8em}}c@{\hspace{0.8em}}c@{\hspace{0.5em}}|@{\hspace{0.5em}}c@{\hspace{0.8em}}c@{\hspace{0.8em}}c@{\hspace{0.8em}}c@{\hspace{0.8em}}c@{\hspace{0.8em}}]}
t=1,l=1 & \:\;\ldots & D_{1,1}b^1_1 & \ldots & D_{1,1}b^1_2 & \ldots & \ldots & Db^1_1& \ldots & Db^1_2 & \ldots & &&&& \\
t=1,l=2 & \:\;\ldots & D_{1,1} b^2_1 & \ldots & D_{1,1} b^2_2 & \ldots & \ldots & D b^2_1 & \ldots & D b^2_2 & \ldots & &&&& \\
t=2,l=1 & \:\;\ldots & D_{2,1} b^1_1& \ldots & D_{2,1} b^1_2 & \ldots & &&&& & \ldots & D b^1_1& \ldots  & D b^1_2 & \ldots \; \\
t=2,l=2 & \:\;\ldots & D_{2,1} b^2_1 & \ldots & D_{2,1} b^2_2 & \ldots & &&&& & \ldots & D b^2_1 & \ldots  & D b^2_2 & \ldots \; \\
\end{block}
\end{blockarray} 
\]
\end{example}

We have the following observation on the pattern of the regrouped rigidity matrix.
\begin{observation}
\label{obs:mat_pattern}
In the rigidity matrix $M$ with columns grouped into column groups,
a hyperedge $e_k$ has $m_k(d-s_k)$ rows, each associated with a  multi-hyperedge of $e^k$ in $\hat{H}$. %(it corresponds to $m_k(d-s_k)$ multi-hyperedges of $\hat{E^k}$\slash{}rows of $M$).
In a column group $j$ where $j \le s_k-1$, each row associated with $e_k$
contains $s_k$ nonzero entries at the columns corresponding to vertices of $e_k$.
In a Column group $j$ where $j = s_k-1+t \ge s_k$, 
a row assocated with a multi-hyperedge $e^k_{r,l}$ of $e^k$ is all zero if $r \ne t$;
the remaining $m_k$ rows contains $s_k$ nonzero entries at the columns corresponding to vertices of $e_k$.
%exactly $m_k$ rows are not all zero
% In Column group $j$ with $j \le s-1$, each row has $s_k$ nonzero entries corresponding to vertices of the hyperedge $e^k$. %where row vertex $v^k_{i}$ has the entry $D^k_{t,j} b^{k,l}_i$ in Row $r^k_{t,l}$.
%\item In Column group $s-1+t$, a row $r^k_{t',l}$ is either all zero if $t' \ne t$, 
%or has $s_k$ nonzero entries $D^k b^{k,l}_i$ corresponding to vertices $v^k_i$'s of the hyperedge $e^k$.
%%For rows $r^k_{t',l}$ with $t' = t$, vertex $v^k_i$ has the entry $D^k b^{k,l}_i$.
\end{observation}

A \emph{labeling of $\hat{H}$ compatible with a given map-decomposition} can be obtained as following.
We start from the last column group of $M$ and associate each column group $j$ with a map in the map-decomposition.
For each multi-hyperedge of the map that is a copy of the hyperedge $e^k$, 
we pick a row $r^k_{t,j}$ that is not all zero in Column groups $j$ and label the multi-hyperedge as $e^k_{t,j}$.
By Observation~\ref{obs:mat_pattern}, this is always possible if each map contains at most $m_k$ multi-hyperedges of the same hyperedge $e_k$, which must be true if there exists any labeling of $\hat{H}$ satisfying Condition 2(a) of Theorem~\ref{thm:rigidity_condition}.
%On the other hand, if some map contains more than $m_k$ multi-hyperedges of $e^k$ and a compatible labeling don't exists, 
%because (1) by Condition~2(a), the number of multi-hyperedges in a map $j$ from a hyperedge $e_k$ is $m(j) \le m_k$.
%\todo{(2) Observation of the row pattern of $M$: fix a hyperedge $e_k$. In Column group $j$ where $j \ge s_k$, the hyperedge  has $m_k$ rows that are not all zero, and the non-zero row indices are non-overlapping across column groups}, so we can just pick any $m(j)$ rows from these $m_k$ rows; \todo{for Column group $j$ where $j \le s-1$, all rows of $e_k$ are not all zero,} so we just pick any $m(j)$ rows from the rows that are not assigned to a multi-hyperedge in a previous map. 

In the Laplace expansion
\begin{equation}
\det({M}) = \sum_{\sigma} \left( \pm \prod_j \det {M}[R_j^\sigma, C_j] \right) \label{eq:laplace}
\end{equation}
the sum is taken over all partitions $\sigma$ of the rows into $d-1$ subsets
$R^\sigma_1, R^\sigma_2,$ $\ldots,$ $R^\sigma_j,$ $\ldots,$ $R^\sigma_{d-1}$, each of size $|V|$.
In other words, each summation term of~\eqref{eq:laplace} contains $|V|$ rows $R_j^\sigma$ from each column group $C_j$.
Observe that for any submatrix ${M}[R^\sigma_{j}, C_{j}]$, each row has a common coefficient $D^k_{t,j}$, so
\[
\det({M}[R^\sigma_{j}, C_{j}]) = \left(\prod_{r^k_{t,j} \in R^\sigma_j} D^k_{t,j} \right) \det(M'[R^\sigma_{j}, C_{j}])
\]
where each row of $M'[R^\sigma_{j}, C_{j}]$ is either all zero, or of the pattern
\begin{equation}
[0,\ldots, 0, b^{k,l}_1, b^{k,l}_2, 0, \ldots ,  b^{k,l}_{s_k}, 0, \ldots, 0] \label{eq:submat_rowpattern}
\end{equation}
with non-zero entries only at the $s_k$ indices corresponding to $v^k_i \in e_k$.

For a fixed $\sigma$, we refer to a submatrix  ${M}[R^\sigma_{j}, C_{j}]$ simply as $M_j$.

\begin{example}
Figure \ref{fig:d3_rigid} shows a pinned subspace-incidence system in $d=3$ with 4 vertices and 5 hyperedges, 
where $e_1 = \{v_1\}, e_2 = \{v_2\}, e_3 = \{v_1, v_3\}, e_4 = \{v_2, v_4\}, e_5=\{v_3,v_4\}$, 
and $m_k = 1$ for all $1 \le k \le 5$ except that $m_5 = 2$.
Figure \ref{fig:d3_rigid_map} gives a map-decomposition of the multi-hypergraph $\hat{H}$ of (a).
The labeling of multi-hyperedges is given in the  regrouped rigidity matrix  \eqref{eq:map_decomp}, 
where the colored rows inside the column groups constitute the submatrices $M^\sigma_j$ in the summation term of the Laplace decomposition corresponding to the map-decomposition.
The system is generically minimally rigid, as the map-decomposition and labeling of $\hat{H}$ satisfies the conditions of Theorem~\ref{thm:rigidity_condition}.
\begin{figure}
\centering
\begin{subfigure}{.55\linewidth}
  \centering
  \includegraphics[width=.95\linewidth]{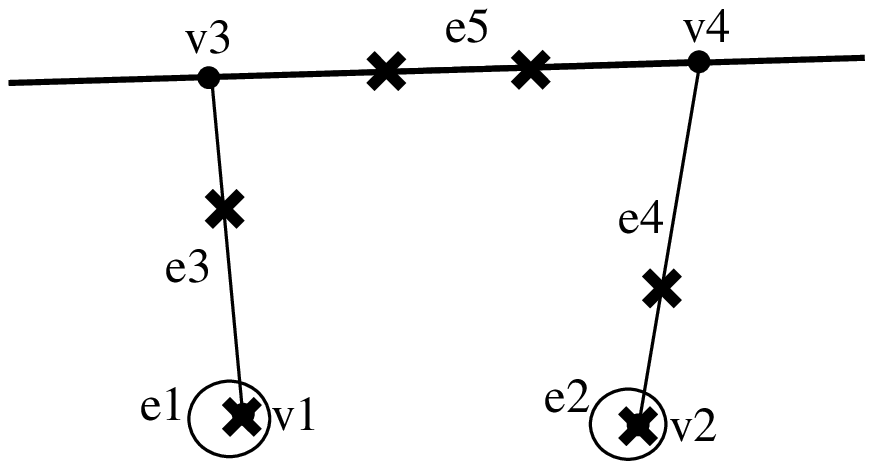}
  \caption{}
  \label{fig:d3_rigid}
\end{subfigure}%
\begin{subfigure}{.45\linewidth}
  \centering
  \includegraphics[width=.8\linewidth]{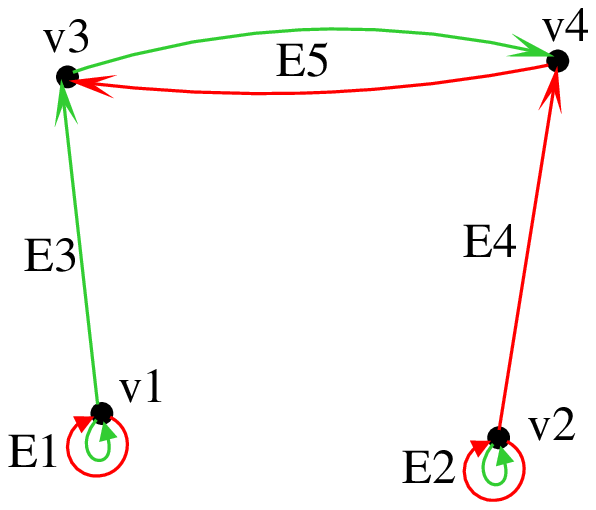}
  \caption{}
  \label{fig:d3_rigid_map}
\end{subfigure}
\caption{(a) A minimally rigid pinned subspace-incidence system in $d=3$.  
(b) A map-decomposition of the multi-hypergraph of the system in (a), where multi-hyperedges with different colors are in different maps, and the tail vertex of each multi-hyperedge is pointed to by an arrow.
}
\end{figure}

\begin{equation*}
\def\arraystretch{1.4}
\begin{blockarray}{c@{\hspace{1em}}cccc|cccc}
  &  v_{1,1} & v_{2,1} & v_{3,1} & v_{4,1} & v_{1,2} & v_{2,2} & v_{3,2} & v_{4,2} \\
\begin{block}{c@{\hspace{1em}}(cccc|cccc)}
\rowcolor{red!20}
e^1_{1,1} & \tikzmarkin[hor=style green]{el} \mathbf{D^1 b^1} &   &   & \hfill\tikzmarkend{el} &   &   \\ 
e^1_{2,1} &   &     &  &  & \tikzmarkin[hor=style orange]{fl} \mathbf{ D^1 b^1 } & & & \hfill \tikzmarkend{fl} \\
e^2_{1,1} & \tikzmarkin[hor=style green]{e2} \hfill\hfill & \mathbf{D^2 b^2} & & \hfill \tikzmarkend{e2} &\\
e^2_{2,1} &  &  & & &  \tikzmarkin[hor=style orange]{f2} \hfill\hfill & \mathbf{D^2b^2} & & \hfill \tikzmarkend{f2}  \\
e^3_{1,1} & \tikzmarkin[hor=style green]{e3} D^3_{1} b^{3}_1 && \mathbf{D^3_{1} b^{3}_2} & \hfill \tikzmarkend{e3} & D^3 b^{3}_1 && D^3b^{3}_2\\
e^4_{1,1} & & D^4_{1} b^{4}_1 && D^4_{1} b^{4}_2 &  \tikzmarkin[hor=style orange]{f3}\hfill\hfill &  D^4b^{4}_1 && \mathbf{D^4b^{4}_2} \hfill \tikzmarkend{f3} \\ 
e^5_{1,1} &  \tikzmarkin[hor=style green]{e4} \hfill\hfill && D^5_1 b^{5,1}_1 & \mathbf{D^5_1 b^{5,1}_2} \tikzmarkend{e4} & && D^5_2 b^{5,1}_1 & D^5_2 b^{5,1}_2  \\
e^5_{1,2} &&& D^5_1 b^{5,2}_1 & D^5_1 b^{5,2}_2 & \tikzmarkin[hor=style orange]{f4}\hfill\hfill && \mathbf{D^5_2 b^{5,2}_1} & D^5_2 b^{5,2}_2 \tikzmarkend{f4} \\ 
\end{block}
\end{blockarray} 
\addtocounter{equation}{1}\tag{\theequation} \label{eq:map_decomp}
\end{equation*}
\end{example}

\begin{proof}[Main Theorem]
First we show the only if direction. 
For a generically minimally rigid pinned subspace-incidence framework, the rigidity matrix $M$ is generically full rank,
so there exists at least one summation term $\sigma$ in \eqref{eq:laplace} where each 
submatrix ${M}_j$ is generically full rank.
As the submatrices don't have any overlapping rows with each other,
we %\todo{divide each row $r^k_{t,l}$ by $D^k_{t,j}$ where $r^k_{t,l} \in R^\sigma_j$, then }  
can perform row elimination on $M$ to obtain a matrix $N$ with the same rank, %convert it to $N$, where all 
where all submatrices ${M}_j$ are simultaneously 
converted to a \emph{permuted reduced row echelon form} ${N}_j$,
where each row in $N_j$ has exactly one non-zero entry $\beta^j_i$ at a unique Column $i$, 
%and different rows have their non-zero entries on different columns.
%(i.e.\ 
In other words, all ${N}_j$'s can be converted simultaneously to reduced row echelon form
by multiplying a permutation matrix on the left of $N$.
%each row contains exactly one non-zero entry, all in different columns. 
%Column group based row elimination gives a non-zero entry for each row, at different columns, inside each $N$. 
%So after elimination the only non-zero entries correspond to the tails for each multi-hyperedge.
%
Now we can obtain a map-decomposition of $\hat{H}$
by letting each map $j$ contain multi-hyperedges corresponding to rows of the submatrix $N_j$,
and assigning each multi-hyperedge in map $j$  the vertex $i$ corresponding to the non-zero entry $\beta^j_i$ in the associated row in ${N}_j$ as tail.
In addition, such a map-decomposition must satisfy Condition 2 in Theorem~\ref{thm:rigidity_condition}:

Condition 2(a): assume two multi-hyperedges $e^k_{t_1,l}$ and $e^k_{t_2,l}$ are in the same map $j$,
i.e.\ the rows corresponding to these two edges are included in the same submatrix $M_j$. 
If $j > s-1$, one of these rows must be all-zero in $M_j$ by Observation~\ref{obs:mat_pattern}, contradicting the condition that $M_j$ is full rank. 
If $j \le s-1$, both of these rows in $M_j$ will be a multiple of the same row vector \eqref{eq:submat_rowpattern},
contradicting the condition that $M_j$ is full rank. 
%Observe from the column group, we know that if they are in the same map $j$, $j \le s-1$. 
%Clearly the abovementioned specification is not possible, 
%and Map $j$ will not have full rank for any specificaiton as one row is a multiple of another.

Condition 2(b): 
%TODO{This argument is not valid when $j_1 \ne j_2$! We need to find a way of elimination / a way to switch tails etc.}
note that the rows  in $M$ corresponding to multi-hyperedges $e^k_{t,l_1}$ and $e^k_{t,l_2}$ have the exactly the same pattern except for different values of $b$'s.
If $e^k_{t,l_1}$ has vertex $i$ as  tail, after the row elimination, the column containing $b^{k,l_1}_{i}$ will become the only non-zero entry of column $i$ for $N_{j_1}$, 
while the column containing $b^{k,l_2}_{i}$ will become zero in all column groups, 
thus $i$ cannot be assigned as tail for $e^k_{t,l_2}$.

%Assume two multi-hyperedges $e^k_{t,l_1}$ and $e^k_{t,l_2}$ are assigned the same tail $v^k_i$ in the map-decomposition.
%In other words, in our row elimination $b^{k,l_1}_{i}$ became the only non-zero entry of column $i$ for $N_{j_1}$,
%and $b^{k,l2}_{i}$ became the only non-zero entry of column $i$ in $N_{j_2}$.
%However, %\todo{As can be observed from the row pattern (?)}, 
%the rows corresponding to these two edges in $M$ have the same pattern except for different values of $b$'s. 
%So in $M_{j_1}$, row 2 has $b^{k,l_2}_{i}$ in column $i$, which became zero after the elimination. 
%Such an elimination will also make the corresponding $b^{k,l_2}_{i}$ in $M_{j_2}$ zero, contradiction.
%%In the reduced matrix $N$, these two rows will have nonzero entries ${\beta_1}^{j_1}_i$ and ${\beta_2}^{j_2}_i$
%%with the same column $i$ in their respective submatrices ${N}[R_{j_1}^\sigma, C_{j_1}]$ and ${N}[R_{j_2}^\sigma, C_{j_2}]$.

%Condition 2(c) is obvious from Observation~\ref{obs:mat_pattern}.

%\todo{(c) is obvious from the observation of the row pattern: for a hyperedge $e_k$, 
%except for the first $s_k - 1$ column groups, 
%any Column group $j$ contains exactly $m_k$ rows that are non-zero, and all those rows have the same index $t$.}

%todo{I think we do need an example for $N$ after row elimination / Maybe notation of entries? }

\medskip

Next we show the if direction, that the conditions of Theorem~\ref{thm:rigidity_condition} imply infinitesimal rigidity. 

Given labeled multi-hypergraph $\hat{H}$ with a map-decomposition satisfying the conditions of Theorem~\ref{thm:rigidity_condition}, 
we can obtain  summation term $\sigma$ in the Laplace decomposition~\eqref{eq:laplace} 
according to the labeling of $\hat{H}$, where
%we can obtain a generically non-zero summation term $\sigma$ in the Laplace decomposition~\eqref{eq:laplace}.
%
%For each map in the map-decomposition,
%associate it with a Column group $j$ where all rows corresponding to multi-hyperedges in the map are not all zero, 
%and let the 
each submatrix $M_j$ contain all rows corresponding to the map associated with Column group $j$.
%This is always possible and gives a Laplace decomposition because of Condition 2 of Theorem~\ref{thm:rigidity_condition} and Observation~\ref{obs:mat_pattern}.
%(1) by Condition~2(a), each map contains at most $m_k$ multi-hyperedges $e^k_{t,l}$ from $e_k$ with different $l$'s,
%(2) by Condition~2(c), at most $s_k-1$ maps can contain multi-hyperedges with different $t$ and $l$, 

We first show that each submatrix $M_j$ is generically full rank. % by giving a specialization.
According to the definition of a map-graph, 
the function $\tau:\hat{E} \rightarrow V$ assigning a tail vertex to each multi-hyperedge  is a one-to-one correspondence.
We perform symbolic row elimination of the matrix $M$ to simultaneously convert each $M_j$ to its permuted reduced row echelon form $N_j$, where for each row of $N_j$, all entries are zero except for the entry $\beta^k_{t,l}$ corresponding to the vertex $\tau(e^k_{t,l})$, which is a polynomial in  $b^{k,l}_i$'s in the submatrix $M_j$.
Since $M_j$ cannot contain two rows with the same $k$ and $l$ by Condition 2(a), the $b^{k,l}_i$'s in different rows of a same map are independent of each other, $\beta^k_{t,l} \ne 0$ under a generic specialization of $b^{k,l}_i$.
%Since the map-decomposition satisfies Condition 2 of Theorem \ref{thm:rigidity_condition},
%each term $b^{k,l}_i$ appears only once in each submatrix $M_j$.
%{the condition does not hold that multi-hyperedges with same $t$ and different $l$ have the same tail!}
%we can use the specialization where $b^{k,l}_i$ is $1$ if  $v^k_i = \tau(e^k_{t_l})$, and is 0 otherwise.
%For a row of $M_j$ associated with $e^k_{t,l}$, we use the specialization where the entry $b^{k,l}_i$ corresponding to the vertex $\tau(e^k_{t,l})$ is $1$ and the remaining entries are all zero.
%
%Under any generic specialization of $b^{k,l}_i$'s, 
Since each row of $N_j$ has exactly one nonzero entry and the nonzero entries from different rows are on different columns, the $|V| \times |V|$ matrix $N_j$ is clearly full rank. Thus $M_j$ must also be generically full rank.

We conclude that 
\begin{equation}
\det({M}) = \sum_{\sigma} \left( \pm \prod_j \Bigg( \Big(\prod_{r^k_{t,j} \in R^\sigma_j} D^k_{t,j} \Big) \det {M}'[R_j^\sigma, C_j] \Bigg) \right) \label{eq:pure_cond}
\end{equation}
where the sum is taken over all $\sigma$ corresponding to a map-decomposition of $\hat{H}$.
Generically, the summation terms of the sum \eqref{eq:pure_cond} do not cancel with each other, 
since $\det(M'[R_j^\sigma,C_j])$ are independent of the multi-linear coefficients $\prod_{r^k_{t,j} \in R^\sigma_j} D^k_{t,j}$, 
and any two rows of $M$ are independent by Condition 2(b).
%TODO I am not sure about this!
This implies that %the generic rank of $\hat{M}$ is $(d-1)m$, 
$\hat{M}$ is generically full rank.

The polynomial \eqref{eq:pure_cond} 
%Taking the terms of \eqref{eq:laplace} that correspond to map-decompositions
%and substituting the values of  $\det(M'[R_j^\sigma,C_j])$ \todo{from Lemma \ref{lem:map_det}} 
gives the pure condition for genericity.	
In particular, 
when there is a subgraph $(V',E')$ with $|V'| < d$ and $\sum_{e_k \in E'} m_k > |V'|$, 
the pure condition vanishes and the system won't be minimally rigid: see Example~\ref{ex:pure_cond}.

\end{proof}

\subsubsection{Pure condition}

%%TODO should the pure condition be here?
%One thing to be noticed is the validity of Theorem~\ref{thm:rigidity_condition} requires the framework
%to avoid certain polynomial called \emph{pure condition} in addition of being generic as in Definition~\ref{def:genericity}
%(see Equation \eqref{eq:pure_condition}).
The pure condition \eqref{eq:pure_cond} obtained in the proof of Theorem~\ref{thm:rigidity_condition} characterizes the badly behaved cases that break the combinatorial characterization of infinitesimal rigidity. 
However, the geometric meaning of the pure condition is not completely clear. 
One particular condition not captured by Theorem~\ref{thm:rigidity_condition} but enforced by the pure condition is that 
there cannot exists a subgraph $(V', E')$ of $H$ with $|V'| < d$ such that $\sum_{e_k \in E'} m_k > |V'|$,
otherwise simple counterexamples can be constructed to the characterization of the main theorem. 
An immediately consequence is that for any hyperedge $e^k$, the dimension $m_k$ of its associated pin must be less than or equal to its cardinality $s_k$.

\begin{example}
\label{ex:pure_cond}

\begin{figure}
\centering
\begin{subfigure}{.5\linewidth}
  \centering
  \includegraphics[width=.95\linewidth]{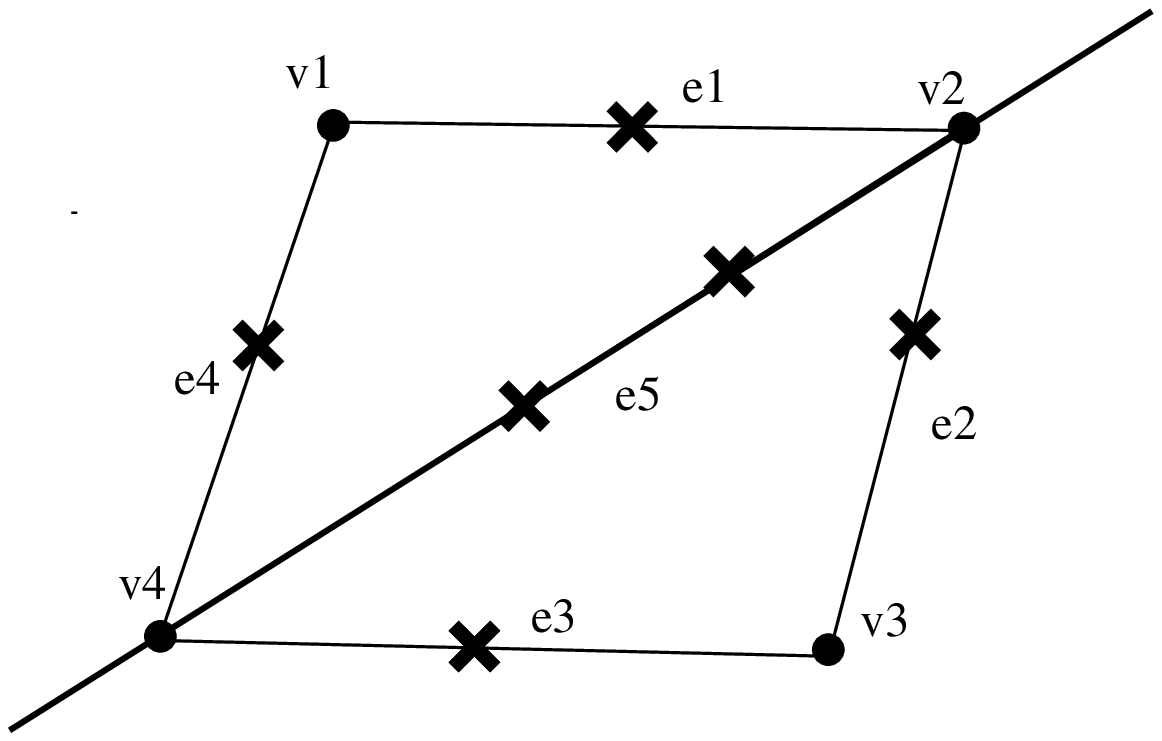}
  \caption{}
  \label{fig:counter_sys}
\end{subfigure}%
\begin{subfigure}{.5\linewidth}
  \centering
  \includegraphics[width=.8\linewidth]{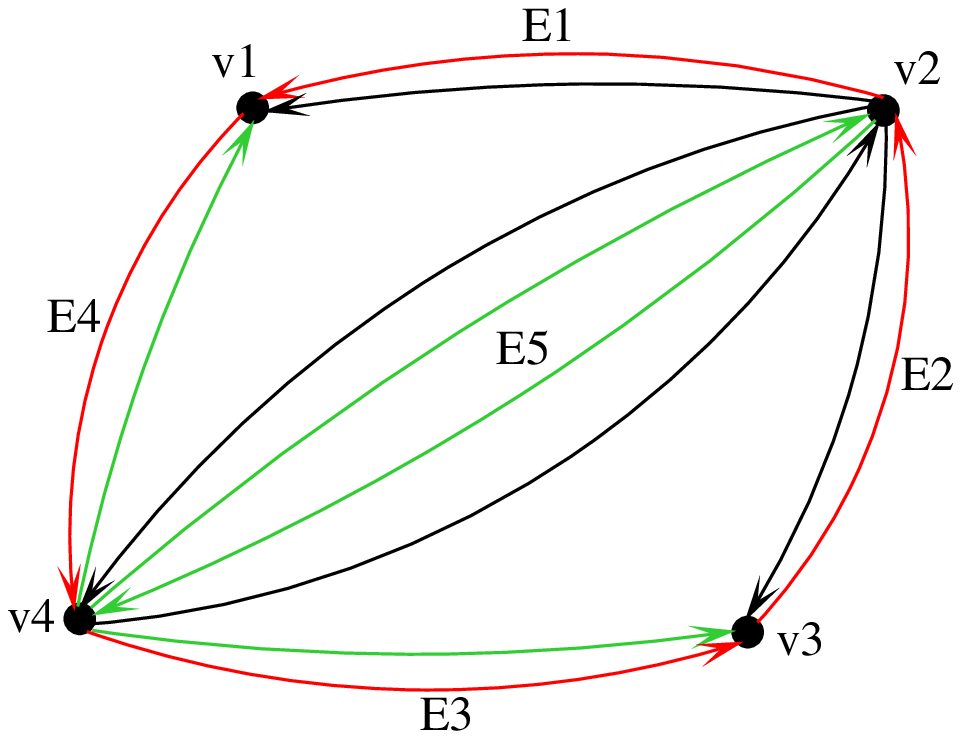}
  \caption{}
  \label{fig:counter_map}
\end{subfigure}
\caption{(a) A pinned subspace-incidence system in $d=4$.  
(b) A map-decomposition of the multi-hypergraph of the system in (a), where multi-hyperedges with different colors are in different maps, and the tail vertex of each multi-hyperedge is pointed to by an arrow.
}
\end{figure}

Figure \ref{fig:counter_sys} shows a  pinned subspace-incidence system in $d=4$ with 4 vertices and 5 hyperedges, where 
$m_k$ is $2$ for $k=5$ and is $1$ otherwise. 
A map-decomposition of the multi-hypergraph $\hat{H}$ of the system is given in Figure~\ref{fig:counter_map}, and we can easily find a labeling of $\hat{H}$ satisfying conditions in Theorem~\ref{thm:rigidity_condition}.
However, the system is not minimally rigid, as generically the pin $x_1$  will not fall on the plane spanned by pins $x_4$ and $x_5$. 
Note that the sub-hypergraph $(V', E')$ spanned by vertices $v_1, v_2, v_4$ violates the pure condition 
as $\sum_{e_k \in E'} m_k = m_1 + m_4 + m_5 = 4 > |V'| = 3$.

\end{example}

\section{Conclusion}
In this paper, we studied the pinned subspace-incidence, a class of incidence geometric constraint system with applications in  dictionary learning and biomaterial modeling. 
We extend our results in \cite{sitharam2014incidence} and
obtain a combinatorial characterization of  minimal rigidity for general pinned subspace-incidence systems
with non-uniform underlying hypergraphs and pins being subspaces with arbitrary dimensions. 

As future work, we plan to extend the underlying group of the pinned subspace-constraint system, 
i.e.\ consider two frameworks to be congruent if the point realization and pin set of one can be obtained from the other under the action of a certain group, for example the projective group.
Another possible direction is to apply  Cayley factorization \cite{farre2013special} to find geometric interpretations of the pure conditions.

\bibliographystyle{plain}
\bibliography{pinned}

\appendix

\section{Proof of Lemma \ref{lem:generic}}

\begin{proof}
%The proof of Lemma~\ref{lem:generic} follows the approach taken by traditional combinatorial rigidity \cite{asimow1978rigidity}.
%%TODO genericity proof
%\todo{
%1. If a framework is infinitesimal rigid, then the corresponding system is rigid. --> inf flex?\\
%2. If a system is rigid, then a generic framework is inf rigid. --> implicit func thm?
%}
First we show that if a framework $(H, X, p)$ is regular, infinitesimal rigidity implies rigidity.
Consider the polynomial system $\algesystem$ of equations. % \eqref{eq:system}, . 
The Implicit Function Theorem states that 
there exists a function $g$, such that $p=g(X)$ on some open interval, if and only if the rigidity matrix $M$ has full rank. 
Therefore, if the framework is infinitesimally rigid, the solutions to the algebraic system are isolated points (otherwise $g$ could not be explicit). 
Since the algebraic system contains finitely many components, there are only finitely many such solution 
and each solution is a $0$ dimensional point. This implies that the total number of solutions is finite,
 which is the definition of rigidity.

To show that generic rigidity implies generic infinitesimal rigidity, we take the contrapositive: if a generic framework is not infinitesimally rigid,
we show that there is a finite flex. 
If $\framework$ is not infinitesimally rigid,
then the rank $r$ of the rigidity matrix ${M}$ is less than $(d-1)|V|$. %the Jacobian $\jacobian$ is less than $2m$. 
Let $E^*$ be a set of edges in $H$ such that $|E^*|=r$ and the corresponding rows in $M$ are all independent.
In ${M}[E^*, \Cdot]$, we can find $r$ independent columns. 
%There are $r$ independent rows as well. 
Let $p^*$ be the components of $p$ corresponding to those $r$ independent columns and $p^{*\perp}$ be the remaining components.
The $r$-by-$r$ submatrix ${M}[E^*, p^*]$, made up of the corresponding independent rows and columns, is invertible. 
Then, by the Implicit Function Theorem, in a neighborhood of $p$ there exists a continuous and differentiable function $g$ such that $p^*=g(p^{*\perp})$.
This identifies $p'$, whose components are $p^*$ and the level set of $g$ corresponding to $p^*$, such that $(H,X)(p')=0$. The level set
defines the finite flexing of the framework. Therefore the system is not rigid.
\end{proof}

\end{document}